\documentclass[aps,prd,preprint,tightenlines,superscriptaddress,
amsfonts,amssymb,amsmath,byrevtex,showpacs,nofootinbib]{revtex4-1}
\usepackage[english]{babel}
\usepackage[applemac]{inputenc}
\usepackage[T1]{fontenc}
\usepackage{latexsym}
\usepackage{graphicx}
\usepackage{geometry}
\usepackage{epstopdf}
\usepackage{subfig}
\usepackage{color}

\usepackage{tikz}

\usepackage{xcolor}
\usepackage{pdfsync}
\usepackage{fancyhdr}
\usepackage{makeidx}
\usepackage[breaklinks,colorlinks,backref]{hyperref}
\hypersetup{
    colorlinks, 
    linktoc=all, 
    linkcolor=red,  
  citecolor=hyptxt,
  urlcolor=blue}
  \hypersetup{
  citebordercolor=,
  filebordercolor=green,
  linkbordercolor=blue
}
\definecolor{hyptxt}{rgb}{0.7, 0.4, 0.9}
\newcommand{\bea}{\begin{eqnarray}}
\newcommand{\eea}{\end{eqnarray}}

\newcommand{\be}{\begin{equation}}
\newcommand{\ee}{\end{equation}}

\newcommand{\ket}[1]{|\kern.3ex#1\kern.3ex\rangle}
\newcommand{\bra}[1]{\langle\kern.3ex #1 \kern.3ex|}
\newcommand{\scalar}[2]{\langle\kern.3ex #1 \kern.3ex|\kern.3ex#2\kern.3ex\rangle}
\newcommand{\norm}[1]{\|\kern.3ex#1\kern.3ex \|}
\newcommand{\dd}{\textnormal{d}}

\newcommand{\CR}{\mathcal{R}}

\newcommand{\CL}{\mathcal{L}}
\newcommand{\CS}{\mathcal{S}}

\def\pt{\partial_{\tau}}
\def\pr{\partial_{\rho}}

\begin{document}

\title{Hamiltonian formulation of dust cloud collapse}

\author{Nick Kwidzinski}
\email{nk@thp.uni-koeln.de}\affiliation{Institute for Theoretical Physics,
University of Cologne, Z\"{u}lpicher Strasse 77, 50937 K\"{o}ln, Germany}

\author{Daniele Malafarina}
\email{daniele.malafarina@nu.edu.kz}
\affiliation{Department of Physics, Nazarbayev University, 53 Kabanbay Batyr, 010000 Nur-Sultan, Kazakhstan}

\author{Jan J. Ostrowski}
\email{Jan.Jakub.Ostrowski@ncbj.gov.pl}
\affiliation{Department of Fundamental Research, National Centre for Nuclear
  Research, Pasteura 7, 02-093 Warszawa, Poland}

 \author{W{\l}odzimierz Piechocki} \email{wlodzimierz.piechocki@ncbj.gov.pl}
\affiliation{Department of Fundamental Research, National Centre for Nuclear
  Research, Pasteura 7, 02-093 Warszawa, Poland}

\author{Tim Schmitz} \email{tschmitz@thp.uni-koeln.de}\affiliation{Institute for Theoretical Physics,
University of Cologne, Z\"{u}lpicher Strasse 77, 50937 K\"{o}ln, Germany}

\date{\today}

\begin{abstract}
We consider the gravitational collapse of a self-gravitating spherical dust cloud in the Hamiltonian formalism.
We address both homogeneous and inhomogeneous cases.  Our novel derivation of the Hamiltonian of the system is
based on an improved variational principle. It differs
from usual treatments due to the presence of an extra boundary term added to the Hilbert action. As expected,
the standard equations of motion are retrieved. However, differently from other treatments, the total Hamiltonian
obtained with our procedure in the Schwarzschild time-gauge is identical to the total mass of the system as
measured from infinity, as it would be expected. Implications for the quantization of the system are suggested.
\end{abstract}


\maketitle

\tableofcontents

\section{Introduction}\label{intr}

Analytical toy models for gravitational collapse in General Relativity (GR) are of great importance from several
perspectives. Firstly, they provide dynamical scenarios that lead to the formation of black holes from regular
initial data \cite{JM}.
Secondly, they can be used to investigate the mathematical properties of the theory, for example in the Hamiltonian
formalism \cite{HK}.
Thirdly, they provide a simple test-bed to study possible pathways towards the quantization of the gravitational field,
since we expect that the singularities that generically arise in GR should not be present in a fully quantum-gravitational
model (see, e.g. \cite{HKA,Gozdz:2018aai}).

Einstein's equations for collapse of perfect fluids are formally identical to the equations describing the universe
expansion in the Friedmann-Lema\^{i}tre-Robertson-Walker (FLRW) cosmological models.
The simplest and most studied solution of Einstein's equations describing collapse is the so-called Oppenheimer-Snyder-Datt
(OSD) model which describes the complete collapse of a spherical cloud composed of homogeneous pressureless matter
(i.e. dust) \cite{OSD}.
It is well known that OSD collapse in co-moving coordinates leads to the formation of a static black hole in a finite co-moving
time, even though far away observers never see the boundary of the cloud crossing the horizon.

The simplest extension of the OSD solution is the so-called Lema\^{i}tre-Tolman-Bondi (LTB) collapse model, which describes
a spherical cloud composed of inhomogeneous dust \cite{LTB}.
The interest in the LTB solution as a theoretical toy model for collapse resides in the fact that, depending on the radial
dependence of the energy-density at the initial time, the complete collapse may produce a naked singularity, i.e. at the
instant of formation, the central singularity may not be covered by the horizon \cite{Joshi}.

Obtaining the Hilbert action and a global Hamiltonian for an asymptotically flat solution of Einstein's equations is
not a trivial matter since typically integrals diverge at spatial infinity. However, one would expect the Hamiltonian
of the system to be related to the total energy as measured by observers at spatial infinity. In \cite{K} it was shown
that by performing the Legendre transformations on an appropriate finite surface one obtains a quasi-local Hamiltonian
which in turn leads to the correct global Hamiltonian once the surface is shifted to infinity.

In the present article we apply the above idea to the OSD and LTB cases. The same procedure was used in the case of
collapse of a thin shell in \cite{KMM}, where the Hamiltonian for an observer at infinity was found to be equal to the
total mass of the collapsing shell regardless of the equation of state of the matter content. As expected, the equations
of motion obtained from the improved variational principle coincide with the equations of motion obtained by different
procedures. For different derivation of this equality see \cite{DS} and references therein.

The importance of finding the Hamiltonian that correctly describes the energy of the system however appears
when one attempts to quantize the system. It is well known that different quantization procedures, based on different
Hamiltonian formulations, provide different results, making nontrivial the issue of quantizing even the simplest
gravitating systems \cite{HKA}.
The hope is that the results obtained for collapse with the improved variational principle may provide a path
towards a viable quantization of gravitational systems.

The paper is organized as follows: In section \ref{mat} we review the matching between the collapsing dust cloud
and the exterior Schwarzschild spacetime. Section \ref{var} is devoted to the application of the variational
principle developed in \cite{K} to the case of dust collapse.
In section \ref{ham} we obtain the Hamiltonian for dust collapse using the above variational principle.  Finally,
in section \ref{con} we summarize the results of this article and mention the possible implications for quantum gravity.

Throughout the paper we make use of natural units setting $c=G=1$.

\section{Matching dust collapse to Schwarzschild exterior}\label{mat}


The general theory for matching two manifolds across a hyper-surface was developed by Darmois and Israel
\cite{Israel}
and it was applied to spherical symmetry by many authors
\cite{matching}.
In the following we will consider the matching between the collapsing dust and an exterior Schwarzschild metric across a spherical surface.
Cosmological models can be used, to some extent, for describing collapse to a black hole (BH) after implementing the condition that one deals with an isolated object.
In our case this reduces to the problem of matching a finite region of FLRW or LTB spacetime with the Schwarzschild geometry. Therefore, in what follows the interior region is described by a portion of FLRW or LTB that extends up to a finite boundary radius, whereas the exterior region is described by the Schwarzschild line element. The interface between the matter field in the interior and the vacuum exterior is used to match both spacetimes.
In what follows we apply the standard method of matching thus requiring continuity of the first and second fundamental forms across the hypersurface separating the two manifolds (see, e.g. \cite{EP}).

\subsection{Matching conditions for FLRW}

For the interior, we consider the homogeneous collapse of dust
with the line element in co--moving hyperspherical coordinates $\{\tau,\chi,\theta,\phi\}$ given by
\be
\dd s^2_-=-\dd \tau^2+a(\tau)^2\left[ \dd \chi^2+h^2(\chi) \dd \Omega^2\right] \, \ .
\ee
We consider the marginally bound (corresponding to flat cosmological models) and bound (corresponding to closed cosmological models) cases simultaneously.
The unbound (i.e. open) case is somehow less relevant for collapse as it describes a matter cloud with positive initial velocity at spatial infinity.
The function $h(\chi)$ is given by
\begin{equation}
h(\chi)=  \begin{cases}
    \chi \, , & \text{in the flat case (k=0)} \\
    \sin(\chi)\, , & \text{in the closed case (k=+1)} \\
    \sinh(\chi)\, , & \text{in the open case (k=-1)}
    \end{cases} \ .
\end{equation}
The interior is filled with a homogeneous  field of comoving dust particles, that is, the energy-momentum tensor is given by
\be
T^{\mu\nu}_- = \epsilon u^\mu u^\nu \ ,
\quad \text{where}\quad
\boldsymbol{u}= u^\mu \partial_\mu =\partial_\tau \ .
\ee
The conservation of energy-momentum implies that
$\epsilon= \epsilon_0 / a^3$ with $\epsilon_0>0$.
For the exterior, we have the Schwarzschild geometry in Schwarzschild coordinates $\{t,r,\theta,\phi\}$
\be
\dd s^2_+=-\left(1-\frac{2M}{r}\right)\dd t^2+\frac{1}{1-\frac{2M}{r}}\dd r^2+r^2\dd\Omega^2 \, .
\label{Schwarzschild1}
\ee
The matching will be performed on the hypersurface $\Sigma$ defined in parametric form by
$\Phi_-(\tau,\chi)=\chi-\chi_b=0$ in the interior, and $\Phi_+(t,r)=r-\psi(t)=0$ in the exterior. Note that $\chi_b>0$ in the flat case while $0<\chi_b<\pi/2$ in the closed case.
On $\Sigma$ we can consider $\tau=f(t)$. We choose the coordinates    $\{ t ,\theta, \phi \}$ to parameterize the hypersurface.
The metrices on $\Sigma$ are then given by
\begin{align}
\dd s^2_+\big|_\Sigma = &
-\left(1-\frac{2M}{\psi}- \frac{\dot{\psi}^2}{1-\frac{2M}{\psi}}  \right)\dd t^2 + \psi^2 \dd \Omega^2, \\
\dd s^2_-\big|_{\Sigma} = &
-\dot{f}^2 \dd t + a^2 h^2(\chi_b)\dd \Omega^2
\ .
\label{Schwarzschild}
\end{align}
Then, the first matching condition  gives (see App. \ref{general}):
\bea
 \dot{f}^2&=&\left(\frac{d\tau}{dt}\right)^2=1-\frac{2M}{\psi}-\frac{\dot{\psi}^2}{1-\frac{2M}{\psi}} \, ,
\label{eq:metric_matching_1}
  \\
 \psi(t)&=& h(\chi_b) a(\tau(t)) \, ,
 \label{eq:metric_matching_2}
\eea
where with dot we denote partial differentiation with respect to the Schwarzschild time $t$.
We can eliminate $\psi$ from \eqref{eq:metric_matching_1} by using \eqref{eq:metric_matching_2} to obtain
\be
\dot{f}^2=\frac{\left(1-\frac{2M}{h (\chi_b) a }\right)^2}
{1-\frac{2M}{h (\chi_b) a }+ (h(\chi_b) a')^2} \ ,
\label{eq:metric_matching_1+2}
\ee
where the prime denotes differentiation with respect to $\tau$.

The  normal co-vectors $\boldsymbol{n}^\pm = n^\pm_\mu \dd x_\pm^{\mu}$ on both sides read
\be
\boldsymbol{n}^+
=|\dot{f}|^{-1}\left(\dd r -  \dot{\psi}\, \dd t \right)
\ , \quad
\boldsymbol{n}^{-}= a \, \dd \chi \ ,
\ee
where we made use of equation \eqref{eq:metric_matching_1} for brevity.
The nonvanishing components of the second fundamental form and their traces on $\Sigma$ are given by
\begin{equation}
\begin{aligned}
K^+_{tt} &=
-\frac{\dot{f}^2}{\dot{\psi}}\partial_t \left[
\frac{1}{|\dot{f}|}\left(1-\frac{2M}{\psi}\right) \right]
\ ,
\\
K^+_{\theta\theta} &= |\dot{f}|^{-1} (\psi -2 M)
\ ,\quad
K^+_{\phi\phi} = |\dot{f}|^{-1} (\psi -2 M)\sin^2 \theta
\ ,
\\
 K^-_{\theta \theta} &= a h \partial_\chi h \big|_{\chi=\chi_b}
\ ,\quad
  K^-_{\phi \phi} = a \sin^2 (\theta)h \partial_\chi h \big|_{\chi=\chi_b} \\
K^- &= \frac{2}{a h}\partial_\chi h \big|_{\chi=\chi_b}
\ .
\end{aligned}
\label{eq:ExK_M=const}
\end{equation}
The matching condition
\be
[K_{\theta\theta}]=0=[K_{\phi\phi}]
\label{eq:matching_Kphiphi}
\ee then yields that
\be
(a')^2+k = \frac{2M}{h^3(\chi_b) a}  \ ,
\label{eq:matching_Kphiphi_Friedmann}
\ee
where we used the first matching condition, that is, \eqref{eq:metric_matching_1} and \eqref{eq:metric_matching_2}.
The remaining matching condition $K^+_{tt}=0$  can be shown to be satisfied if the other matching conditions \eqref{eq:metric_matching_1}, \eqref{eq:metric_matching_2} and \eqref{eq:matching_Kphiphi} are already imposed.

So far we have three equations that might be used to determine the unknown functions $a$, $\psi$ and $f$. What remains undetermined is the mass parameter $M$ in the exterior region. This can be fixed by imposing the Friedmann equation
\begin{equation}
(a')^2+k=\frac{8\pi}{3}\epsilon a^2 \ .
\end{equation}
Comparing this to equation \eqref{eq:matching_Kphiphi_Friedmann} yields
\be
M = \frac{4\pi}{3} h^3(\chi_b)\epsilon_0  \ .
\ee

Let us now investigate the hyperbolic angle $\mu$ between surfaces $\tau=\text{const.}$ on one side
and surfaces $t=\text{const.}$ on the other. The unit vector orthogonal to the surfaces $\tau=\text{const.}$ on the OSD side is
\be
\label{eq:vector}
m_-^\mu=\{1,0,0,0\}\, ,
\ee
whereas the unit vector orthogonal to the surfaces $t=\text{const.}$ on
the Schwarzschild side reads
\be
m_+^\mu=\left\{\frac{1}{\sqrt{1-2M/r}},0,0,0\right\} \, .
\ee

The hyperbolic angle $\mu$ is defined to be
\be
|\mu|:={\rm arcosh}|g_{\mu\nu}m_+^\mu m_-^\nu| \, .
\ee
Consider the normalized four velocity of the dust particles on the boundary $\Sigma$. In the FLRW coordinates it is given by
\be
\boldsymbol{u}= \partial_\tau = m_-^\mu \partial_\mu = \boldsymbol{m}_-  \ ,
\ee
while in the Schwarzschild one's it is given by
\be
\boldsymbol{u}= \frac{\partial_t + \dot{\psi}\partial_r}{\| \partial_t + \dot{\psi}\partial_r \|} \, .
\ee
Since $\boldsymbol{m}_-=\boldsymbol{u}$ it follows that
\be
\cosh \mu= | \langle \boldsymbol{u} , \boldsymbol{n}  \rangle | = \frac{1-\frac{2M}{\psi}}{\sqrt{
\left(1-\frac{2M}{\psi} \right)^2-\dot{\psi}^2}}  \ .
\label{cosh_alpha}
\ee
Furthermore we will choose the sign of $\mu$ in such a way that it coincides with the sign of $\dot{\psi}$. One can then write
\begin{equation}
    \dot{\psi} = \left(1-\frac{2M}{\psi}\right)\tanh \mu \label{eq:psidot}\, .
\end{equation}

Using the above equation, the matching condition \eqref{eq:matching_Kphiphi}, which is simply the second Friedmann equation, can then be written as
\begin{equation}
    1-\frac{2M}{\psi}=\frac{1-h^2(\chi_b)k}{\cosh^2\mu}\, \label{eq:Friedmann_mu}.
\end{equation}

\subsection{Matching conditions for LTB}

The above derivation can be extended in a straightforward way to the case of inhomogeneous dust. Again we use co-moving coordinates $\{\tau, \rho, \theta, \phi\}$ on the LTB side and Schwarzschild coordinates $\{t,r,\theta,\phi\}$ on the
Schwarzschild side.

The Schwarzschild metric is given by \eqref{Schwarzschild1} while the LTB metric reads
\begin{equation}
\label{ltb-synch}
\dd s_{-}^2 = -\dd \tau^2 + \frac{(\pr R)^2}{1+2E}\dd \rho^2 + R^2\dd \Omega^2   \;,
\end{equation}
where $R=R(\rho,\tau)$ is the aerial radius and $E=E(\rho)$ is one arbitrary function resulting from the integration of Einstein's equations. Another function that is useful to consider is the Misner-Sharp mass $F(\rho)$ which defines the amount of matter contained within the co-moving shell labelled by $\rho$ \cite{misner}.
$F(\rho)$ is related to the dust density by
\begin{equation}
    4\pi  \epsilon = \frac{\partial_{\rho}F}{R^2\pr R} \, ,\label{eq:eom_ltb}
\end{equation}
from which it is easy to see that it represents the active gravitational mass. Notice that for dust collapse $F$ does not depend on $\tau$, meaning that the amount of matter contained within the co-moving radius $\rho$ is conserved throughout collapse.
Then the system of Einstein's equations is fully solved once one integrates the equation of motion for $R(\tau,\rho)$ which can be given in the form
\begin{equation}
  \label{LTB-eq1}
    \left(\pt R \right)^2 =\frac{2F}{R}+2E \;.
\end{equation}

Solutions to \eqref{LTB-eq1} read (in parametric form):
\begin{eqnarray}
&E<0& \;\;,\;\;R=-\frac{F}{2E}\left(1-\cos \alpha\right) \;\;,\;\;\alpha - \sin \alpha =\frac{\left(-2E\right)^{3/2}}{F}\left(\tau-t_B\right) \, ; \\
&E=0&\;\;,\;\;R=\left(\frac{9}{2}F\left(\tau-t_B\right)^2\right)^{1/3} \, ; \\
&E>0& \;\;,\;\;R=\frac{F}{2E}\left(\cosh \alpha -1\right)\;\;,\;\;\sinh \alpha -\alpha = \frac{\left(2E\right)^{3/2}}{F}\left(\tau-t_B\right) \, ,
\end{eqnarray}
where $\alpha$ is an auxiliary angle and $t_B$ is another function of $\rho$, which in cosmology is called the big bang time.
Notice that in general, Schwarzschild exterior and LTB interior may define one LTB spacetime as the Schwarzshild metric belongs to the LTB family.

The matching hypersurface $\Sigma$ has the topology of $\mathbb{S}^2 \times \mathbb{R}$.
Similarly to the OSD case, we parameterize this hypersurface in the following way: $\Phi_-(\tau, \rho)=\rho-\rho_b=0$ from the interior, and $\Phi_+(t,r) = r-\psi(t)=0$ from the exterior.
It follows that
\begin{equation}
\dd r = \dot{\psi}\;\dd t \, ,
\end{equation}
and by choosing $\{t,\theta,\phi\}$ with $\tau = f(t)$
as a coordinates on the hypersurface we get the line elements as
\begin{eqnarray}
\dd s_+\big\vert_{\Sigma} &=& -\left(1-\frac{2M}{\psi} - \frac{\dot{\psi}^2}{1-\frac{2M}{\psi}}\right)\dd t^2 + \psi^2\dd \Omega^2 \, ,\nonumber \\
\dd s_-\big\vert_{\Sigma} &=& -\dot{f}^2 \dd t^2 + R_b^2 \dd \Omega^2 \;\;,\;\;R_b = R(\rho_b,t)\, .
\end{eqnarray}
Again, we have the following conditions for continuity of the metric, i.e. the first matching conditions:
\begin{eqnarray}
  \dot{f}^2 &=&  1-\frac{2M}{\psi} - \frac{\dot{\psi}^2}{1-\frac{2M}{\psi}} \, ,\\
  \psi &=& R_b \;\;,\;\; \dot{\psi}(t) = \frac{\partial \tau}{\partial t}\;\pt R_b = \dot{f}\;\pt R_b\, .
\end{eqnarray}
In analogy with equation \eqref{eq:metric_matching_1+2}, combining the above equations gives
\begin{equation}
  \label{eq:match}
  \dot{f}^2 = \frac{\left(1-\frac{2M}{R_b}\right)^2}{1-\frac{2M}{R_b}+\left(\pt R_b\right)^2}\, .
\end{equation}
Normal vectors to the boundary hypersurface read
\begin{equation}
{\bf n}^+ = |\dot{f}|^{-1}\left(\dd r -\dot{\psi}\dd t\right)  , \;\;\;{\bf n}^- = \left|\pr R_b/\sqrt{1+2E_b}\right|\dd \rho \, .
\end{equation}
We can calculate the extrinsic curvature of the boundary surface as:
\begin{eqnarray}
  K^+_{tt} &=&-\frac{\dot{f}^2}{\dot{\psi}}\partial_t \left[\frac{1}{|\dot{f}|}\left(1-\frac{2M}{\psi}\right)\right]\, ; \nonumber \\
  K^+_{\theta \theta} &=& |\dot{f}|^{-1}\left(\psi - 2M\right) \;\;,\;\;K^-_{\theta \theta} = R_b\sqrt{1+2E_b} \, ;\nonumber \\
  K^+_{\phi \phi} &=& K^+_{\theta \theta} \sin^2(\theta) \;\;,\;\; K^-_{\phi \phi} = K^-_{\theta \theta} \sin^2(\theta)\, .
\end{eqnarray}
In addition, the trace of the extrinsic curvature reads
\begin{equation}
K^- = \frac{2\sqrt{1+2E_b}}{R_b}\, .
  \end{equation}
The second matching conditions then reduce to
\begin{equation}
K^+_{\theta \theta} = K^-_{\theta \theta} \;\;, \;\;|\dot{f}|^{-1}\left(\psi - 2M\right) = R_b\sqrt{1+2E_b}
\, ,
  \end{equation}
and using \eqref{eq:match} we obtain
\begin{equation}
\left(\pt R_b\right)^2=\frac{2M}{R_b}+2E_b \, ,
  \end{equation}
which is equivalent to \eqref{LTB-eq1} provided that $M=F_b$.

Finally, performing calculations analogously to \eqref{eq:vector}-\eqref{cosh_alpha},
we can express the field equation \eqref{LTB-eq1} in terms of hyperbolic angle (with no explicit time derivative) as follows
\begin{equation}
1-\frac{2M}{\psi} = \frac{1+2E_b}{\cosh^2\mu} \label{eq:eom_LTB} \, .
  \end{equation}

\section{Variational principle}\label{var}

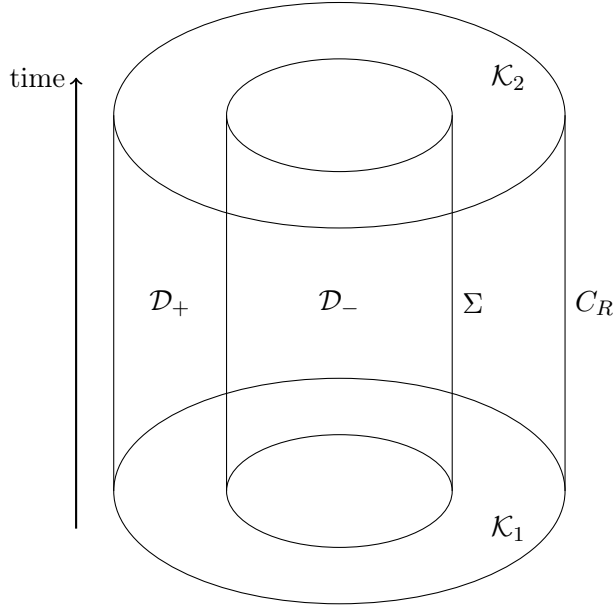
\begin{figure}
	\centering
	\begin{tikzpicture}
		\draw (0,0) ellipse (3cm and 1.5cm);
		\draw (0,0) ellipse (1.5cm and 0.75cm);
		\draw (0,5cm) ellipse (3cm and 1.5cm);
		\draw (0,5cm) ellipse (1.5cm and 0.75cm);
		\draw (-3cm,0) -- (-3cm,5cm);
		\draw (3cm,0) -- (3cm,5cm) node [midway,right] {$C_R$};
		\draw (-1.5cm,0) -- (-1.5cm,5cm);
		\draw (1.5cm,0) -- (1.5cm,5cm) node [midway,right] {$\Sigma$};
		\draw (0,2.5cm) node {$\mathcal{D}_-$};
		\draw (-2.25cm,2.5cm) node {$\mathcal{D}_+$};
		\draw (2.25cm,5.5cm) node {$\mathcal{K}_2$};
		\draw (2.25cm,-0.5cm) node {$\mathcal{K}_1$};
		\draw[thick, ->] (-3.5cm,-0.5cm) -- (-3.5cm,5.5cm) node [at end, left] {time};
	\end{tikzpicture}
	\caption{Total domain of integration $\mathcal{D}$, consisting of Friedmann (and LTB) interior $\mathcal{D}_-$ and Schwarzschild exterior $\mathcal{D}_+$, with $\Sigma$ as the matching surface. As boundaries we have a surface of constant Schwarzschild radius $C_R$, where we take $R\to\infty$ at the end, and two surfaces of constant time $\mathcal{K}_{1/2}$. Note that $\mathcal{K}_{1/2}$ are not smooth where they overlap $\Sigma$, since in the interior we use dust proper time, and in the exterior Schwarzschild Killing time.}
	\label{fig:domains}
\end{figure}

Following \cite{KMM} the full action for our model contains the following contributions
\begin{equation}
S=\int_{\mathcal{D}_-}\mathcal{L}_{\text{grav}}+
\int_{\mathcal{D}_-}\mathcal{L}_{\text{dust}} +
\int_{\Sigma}\mathcal{L}_{\text{grav}} +
\int_{\mathcal{D}_+}\mathcal{L}_{\text{grav}} +
\int_{\partial\mathcal{D}}\mathcal{L}_{\text{boundary}}
\label{eq:improved_action}
\end{equation}
The domains are sketched in figure \ref{fig:domains}. The matching between Friedmann or LTB interior $\mathcal{D}_-$ and Schwarzschild exterior $\mathcal{D}_+$ is implemented here by a boundary term on the matching hypersurface $\Sigma$, given by the jump in the trace of the ADM momentum on $\Sigma$. The mass of the Schwarzschild exterior is taken to be dynamical, $M=M(t)$. All the necessary Gibbons-Hawking-York (GHY) terms are included in the boundary term $\int_{\partial\mathcal{D}}\mathcal{L}_{\text{boundary}} $. In addition to the GHY terms on the surfaces of constant time $\mathcal{K}_1$ and $\mathcal{K}_2$,   we require   edge terms on the intersection $\Sigma \cap \mathcal{K}$, where $\mathcal{K}:=\mathcal{K}_1 \cup \mathcal{K}_2$. For a discussion of such terms see  \cite{Hayward}. Lastly we require a GHY term on the spacelike boundary of infinite curvature radius $C_R\big|_{R\to\infty}$.

When we have the Lagrangian in place, the dynamical quantities to be varied will be the degrees of freedom left open for the FLRW or LTB interior and Schwarzschild exterior: the rescaled scale factor $\psi(t)$ and the Schwarzschild mass $M(t)$. As usual when applying the variational principle these dynamical quantities are kept fixed at the timelike boundaries $\mathcal{K}_{1/2}$.

We will further work in a gauge fixed picture where we choose the coordinate frames in the interior and exterior as we have done in the last section.
The final result will thus be an actual physical Hamiltonian generating evolution in Schwarzschild Killing time $t$, and not a constraint.

\subsection{Schwarzschild exterior}

Let us first take care of all exterior terms in the action. We start with the boundary terms in $\int_{\partial\mathcal{D}}\mathcal{L}_{\text{boundary}} $, excluding the one for $\mathcal{D}_+ \cap \mathcal{K}_{1/2}$. The latter will be discussed with the respective interiors. First we have the GHY term on $C_R$. For this we need the extrinsic curvature on that surface.
The spacelike normal vector on              $C_R$ is given by
\begin{equation}
\boldsymbol{n} = \sqrt{1-\frac{2M}{R}}\partial_r \, .
\end{equation}
The only non-vanishing component of $n_\mu$ is $n_r=1/\sqrt{1-2M/r} $.
Therefore the extrinsic curvature tensor is given by
\begin{equation}
K_{ij}= -\Gamma^r_{ij}n_r \ , \quad \text{where}\quad i,j=t,\theta,\phi \ .
\end{equation}
The non-vanishing components of the Christoffel symbols $\Gamma^r_{ij}$ are
\begin{equation}
\Gamma^r_{tt}=   \frac{(R-2M)M}{R^3} \   , \quad
\Gamma^r_{\theta \theta} = -(R-2M) \   , \quad
\Gamma^r_{\phi \phi} = -(R-2M)\sin^2 \theta \ .
\end{equation}
Finally we obtain the trace of the ADM momentum as defined in Appendix \ref{general},
\begin{equation}
Q=2(3M-2R)\sin \theta.
\end{equation}
This gives for the corresponding GHY boundary term
\begin{equation}
 -    \frac{1}{16\pi}\int_{C_R}\dd^3 y \ Q
    = R (t_2-t_1)-\frac{3}{2}\int^{t_2}_{t_1} \dd t\, M
\end{equation}
To obtain the final expression for $\int_{C_R}\mathcal{L}_{\text{boundary}}$ we need take the term from above and subtract  the same surface term which we would obtain by embedding the surface into flat spacetime, e.g. by using the metric
\begin{equation}
    \dd s_0^2 =- \left(1-\frac{2M}{R}\right)\dd t^2 + \dd r^2 + r^2 \dd \Omega^2
\end{equation}
and considering the surface $r=R$.
The trace of the  ADM-momentum for this surface is given by
\begin{equation}
    Q_0 = -2 (R-M) \sin \theta + \mathcal{O}(1/R^2)\ .
\end{equation}
Then
\begin{equation}
\int_{C_R}\mathcal{L}_{\text{boundary}}=-\frac{1}{16\pi}
\lim_{R \rightarrow \infty}
\int_{\partial\mathcal{D}} \dd^3 y (Q-Q_0)=-\frac{1}{2}\int^{t_2}_{t_1} \dd t\, M
\ .
\end{equation}

Next we compute the GHY term on surfaces of constant $t=t_{1/2}$ (i.e. $\mathcal{D}_+ \cap \mathcal{K}_{1/2}$). The
timelike normal vector on that surface reads
\begin{equation}
\boldsymbol{n}= \pm\frac{1}{\sqrt{1-\frac{2M}{r}}}\partial_t \, ,
\end{equation}
where we take the upper sign for the surface with constant $t_2$ and the lower sign for the surface with constant $t_1$. The
only non-vanishing component of the co-vector $n_\mu$ is $n_t=\mp\sqrt{1-\frac{2M}{r}} $.
Therefore the extrinsic curvature on $\mathcal{D}_+ \cap \mathcal{K}_{1/2}$ is given by
\begin{equation}
K_{ij}=-\Gamma^t_{ij}n_t \ , \quad \text{where}\quad i,j=r,\theta,\phi \ .
\end{equation}
The only non vanishing component   of the connection is $\Gamma^t_{rr}=\dot{M}r^2/(r-2M)^3$ and therfore the only non vanishing component of the extrinsic curvature is
\begin{equation}
K_{rr} = \pm \left. \frac{\dot{M}r^{3/2}}{\left(r-2M \right)^{5/2}}\right|_{t=t_{1/2}}
\, .
\end{equation}
This gives
\begin{equation}
    Q=\pm2\frac{\dot{M}r^3}{(r-2M)^2} \sin\theta \, .
\end{equation}
The corresponding boundary term is therefore
\begin{equation}
\int_{\mathcal{D}_+ \cap\, \mathcal{K}_{1}}\mathcal{L}_{\text{boundary}}
+\int_{\mathcal{D}_+ \cap\, \mathcal{K}_{2}}\mathcal{L}_{\text{boundary}}
= \frac{1}{2}\left[\int_{\psi(t)}^R \dd r \frac{\dot{M}r^3}{(r-2M)^2} \right]^{t_2}_{t_1} \, .
\label{eq:GHY_K}
\end{equation}
There are joint terms on the intersections between $\Sigma$ with $\mathcal{K}_1$ and $\mathcal{K}_2$.
Following \cite{Hayward} they are given by
\begin{equation}
    \int_{\Sigma\cap\, \mathcal{K}_{1/2}} \mathcal{L}_{\text{boundary}} =  \frac{1}{8\pi} \int \sqrt{\gamma } \dd^2 y\, \text{arsinh}\! \left( n_\mu m^\mu  \right)\, ,
\end{equation}
where $\boldsymbol{n}=1/\sqrt{1-2M/\psi} \dd t$ is the normal covector to   $\mathcal{K}_{1/2}$ on the Schwarzschild side and $\boldsymbol{m}=-(1/\dot{f})\dd x$ is the normal covector to $\Sigma$. We obtain that
\begin{equation}
  \int_{\Sigma\cap\, \mathcal{K}} \mathcal{L}_{\textnormal{boundary}}
  =\frac{1}{2}\left[\psi^2\mu \right]^{t_2}_{t_1} \ .
  \label{eq:edge_term}
\end{equation}

Next, we want to compute the Schwarzschild bulk term
\begin{equation}
\int_{\mathcal{D}_+}\mathcal{L}_{\text{grav}}=\frac{1}{16\pi}\int_{\mathcal{D}_+} \dd^4 x \sqrt{-g_+}\  {}^{(4)}\CR_+ \, .
\end{equation}
The Ricci scalar for the Schwarzschild spacetime with a time dependent mass term $M=M(t)$ is given by
\begin{equation}
{}^{(4)}\CR_+
=\frac{\partial}{\partial t}\frac{2\dot{M}r}{(r-2M)^2} \ .
\end{equation}
Using Leibniz's integral rule we find that
\begin{equation}
\begin{aligned}
\int_{\mathcal{D}_+} \dd^4 x \sqrt{-g_+} \  {}^{(4)}\CR_+
& =
4\pi \int^{t_2}_{t_1} \dd t \int^R_{\psi(t)} \dd r \ r^2 \  {}^{(4)}\CR_+
\\
& =
8\pi\left[\int_{\psi(t)}^R \dd r \frac{\dot{M}r^3}{(r-2M)^2} \right]^{t_2}_{t_1}
+8\pi \int^{t_2}_{t_1} \dd t
\frac{\dot{\psi}\dot{M}\psi^3}{(\psi-2M)^2} \ .
\end{aligned}
\end{equation}
Hence
\begin{equation}
\int_{\mathcal{D}_+}\mathcal{L}_{\text{grav}}=\frac{1}{2}\left[\int_{\psi(t)}^R \dd r \frac{\dot{M}r^3}{(r-2M)^2} \right]^{t_2}_{t_1}
 + \frac{1}{2} \int^{t_2}_{t_1} \dd t
\frac{\dot{\psi}\dot{M}\psi^3}{(\psi-2M)^2}\, .
\label{eq:Schwarzschilde_bulk}
\end{equation}
The first term on the right hand side cancels with the GHY term on $\mathcal{K}\cap \mathcal{D}_+$ given by equation
\eqref{eq:GHY_K}.

\subsection{FLRW interior}

Next,  we want to consider the gravitational and matter terms on the Friedmann side, that is on $\mathcal{D}_-$ including the GHY terms on  $\mathcal{D}_- \cap \mathcal{K}$,

\begin{equation}
    S_\text{interior}=\int_{\mathcal{D}_-}\mathcal{L}_{\text{grav}}+ \int_{\mathcal{D}_-}\mathcal{L}_{\text{dust}} +
\int_{\mathcal{D}_-\cap\,\mathcal{K}}\mathcal{L}_{\text{boundary}} \ .
\end{equation}

The Gibbons-Hawking-York term  $S_{\text{GHY} }$ provides a contribution on the spacelike boundary at constant times on the Friedmann side.
The Ricci scalar on the Friedmann side is given by
\begin{equation}
{}^{(4)}\CR_- = \frac{6}{a^2}\left(a'^2 + a a'' + k \right) \ ,
\end{equation}
and the trace for the extrinsic curvature on $\mathcal{D}_- \cap \mathcal{K}$ is $K=3a'/a$.

According to Kucha\v{r} and Brown \cite{KucharBrownDust} we can model the matter action by
\begin{equation}
\mathcal{L}_{\text{dust}} = -\frac{1}{2}\sqrt{-g} \
\epsilon \left( g_{\mu\nu} u^\mu u^\nu+1 \right) \ ,
\end{equation}
with $\epsilon$ being the dust density and $\boldsymbol{u}=\dd T=T'\dd \tau$ being the four velocity co-vector 
of the dust field. The variable $T(\tau)$ is the  proper time of the dust particles. The dust density $\epsilon$ is not dynamical and serves a Lagrange multiplier, which ensures the normalization of $\boldsymbol{u}$.

The full action now reads
\begin{equation}
S_\text{interior}= \frac{C }{2} \int\limits_{\tau_1}^{\tau_2} \dd \tau \ a^3 \left[-\frac{ a'^2}{a^2} + \frac{k}{a^2}  +  \frac{4\pi }{3} \epsilon\left(T'^2-1  \right) \right] \, ,\\ \label{eq:FLRW_action_dust}
\end{equation}
where we defined the constant $C:=3\int_{0}^{\chi_b}\dd \chi \  h^2(\chi)$.

For the sake of simplicity we first perform the Legendre transform in the matter sector.
The momentum conjugate to the dust proper time reads
\begin{equation}
 p_T = \frac{4 \pi C a^3 \epsilon}{3}T'
\ .
\label{eq:momentum}
\end{equation}
Variation with respect to $\epsilon$ yields the primary constraint $p_\epsilon=\partial L/\partial \epsilon '  = 0$.  The secondary constraint $p_\epsilon'=0$ yields that  $\epsilon=3 p_T/(4\pi C a^3)$. We immediately implement this constraint equation after performing the Legendre transform in $T$. The action then becomes
\begin{equation}
S_\text{interior}= \frac{C }{2}\int\limits_{\tau_1}^{\tau_2} \dd \tau \ \left[   a^3 \left(-\frac{ a'^2}{a^2} + \frac{k}{a^2}
-  \frac{2p_T }{C a^3}
\right)\right] \, .
\end{equation}
Since $p_T$ is a constant of motion we can keep it in the action not as a dynamical quantity but as an external parameter, controlling the dust density in the interior. To this end we have dropped in the above the Liouville term $p_T T'$. Note that matching exterior and interior as done in the last section leads to the identification $M=h^3(\chi_b)p_T/C$, but we will not implement this before variation.

We now switch to the variable $\psi=a/h(\chi_b)$ and to the Schwarzschild time $t$ to get
\begin{equation}
\begin{aligned}
 S_\text{interior} &= \frac{1}{2}\int\limits_{t_1}^{t_2} \dd t \sqrt{1-\frac{2M}{\psi}-\frac{\dot{\psi}^2}{1-\frac{2M}{\psi}}}
 \left(-2p_T + \frac{C k\psi  }{h(\chi_b)}
-\frac{C\psi}{h^3(\chi_b)} \frac{\dot{\psi}^2}{1-\frac{2M}{\psi}-\frac{\dot{\psi}^2}{1-\frac{2M}{\psi}}}
  \right) \\
  &= \frac{1}{2}\int\limits_{t_1}^{t_2} \dd t\sqrt{1-\frac{2M}{\psi}}\left[\frac{-2p_T + \frac{C}{h(\chi_b)}k\psi}{\cosh\mu} - \frac{C}{h^3(\chi_b)} (\psi-2M )\frac{\sinh^2\mu}{\cosh\mu} \right] \ .
    \label{eq:Lagrangian_interior}
  \end{aligned}
\end{equation}

Next, we wish to compute the term on the matching surface $\Sigma$.
We need to compute the jump $[Q]$ in the trace of the ADM momentum on the surface $\Sigma$.
We already have the extrinsic curvature on the Friedmann side. The result on the Schwarzschild side in \eqref{eq:ExK_M=const} changes due to the time dependence of $M$.
For the computation we switch to the coordinate
$x=r-\psi$.
The normal co-vector is then
\begin{equation}
    \boldsymbol{n} = \frac{1}{\dot{f}}\dd x \, .
\end{equation}
$K^+_{\theta \theta}$ and $K^+_{\phi \phi}$  stay the same as in \eqref{eq:ExK_M=const}, but
\begin{equation}
\begin{aligned}
K^+_{tt}  & =
-\frac{\dot{f}^2}{\dot{\psi}}\partial_t \left[
\frac{1}{|\dot{f}|}\left(1-\frac{2M}{\psi}\right) \right]
- \frac{\dot{M}\dot{f}\left[\psi^2\dot{\psi}^2+(\psi-2M)^2 \right]}{\psi\dot{\psi}(\psi-2M)^2}\, .
\end{aligned}
\end{equation}
From this we get
\begin{equation}
[Q] = 4 \sin \theta \,  \left[\frac{\psi^2}{2\dot{f}}K^+_{tt} - \psi +2M +\partial_\chi h(\chi_b)\psi\dot{f}\right] \, .
\end{equation}
Finally
\begin{align}
\int_{\Sigma}\mathcal{L}_{\text{grav}}&=-\frac{1}{16\pi}\int_\Sigma [Q]\dd^3 y \nonumber\\
&=
   \int^{t_2}_{t_1} \dd t
  \left(
-\frac{1}{2}\frac{\dot{\psi}\dot{M}\psi^3}{(\psi-2M)^2}-\frac{1}{2}\psi^2\dot{\mu}+\frac{3M}{2}-\psi +\partial_\chi h(\chi_b)\dot{f}\psi \right) \ ,
\end{align}
where we have used that
\begin{equation}
\frac{\psi^2\ddot{\psi}}{2\dot{f}^2}=\frac{1}{2} \psi^2\dot{\mu} + \frac{\dot{\psi}\psi}{\dot{f}^2\left( 1-\frac{2M}{\psi}\right)}\left( -\dot{M}+\frac{M\dot{\psi}}{\psi}\right).
\end{equation}
The first term cancels with the one from the Schwarzschild bulk \eqref{eq:Schwarzschilde_bulk}, and the second term can be partially integrated. The resulting boundary term on the $t=\text{const.}$ surfaces cancels with the edge \eqref{eq:edge_term}.

Now, combining everything we see that what remains is the Friedmann term \eqref{eq:Lagrangian_interior} and the boundary term on $C_R$ and $\Sigma$ apart from the first term.
Finally the Lagrangian is given by
\begin{multline}
	L_\text{tot} = -\frac{\psi \sqrt{1-\frac{2M}{\psi}}}{\cosh\mu}\left[\frac{C}{2h^3(\chi_b)}\left(1-\frac{2M}{\psi} \right) \sinh^2\mu -\frac{C}{2h(\chi_b)}k + \frac{p_T}{\psi} - \partial_\chi h(\chi_b)\right]+\\+M-\psi+\psi\dot{\psi}\mu \,. \label{eq:Ltot}
\end{multline}
Since this Lagrangian does not depend on $\dot{M}$, one can express the mass as a function of $\psi$ and $\dot{\psi}$ by enforcing the equation of motion\ for $M$,
\begin{equation}
	\frac{\partial L_\text{tot}}{\partial M}=0\,.
\end{equation}
Using
\begin{equation}
	\frac{\partial \mu}{\partial M}= \frac{2}{\psi}\frac{\sinh\mu\cosh\mu}{1-\frac{2M}{\psi}}
\end{equation}
we then find
\begin{equation}
	-\sqrt{1-\frac{2M}{\psi}} \cosh\mu + \frac{C}{2h^3(\chi_b)}\left(1-\frac{2M}{\psi} \right) \sinh^2\mu= -\frac{C}{2h(\chi_b)}k + \frac{p_T}{\psi} - \partial_\chi h(\chi_b)\,. \label{eq:M}
\end{equation}

This expression will now define $M$ as a function of $\psi$ and $\mu$. We can then simplify
\eqref{eq:Ltot} and get
\begin{equation}
	L_\text{tot} = -\frac{C}{h^3(\chi_b)}\psi \left(1-\frac{2M}{\psi} \right)^\frac{3}{2}\frac{\sinh^2\mu}{\cosh\mu} -M+\psi\dot{\psi}\mu \,.\label{eq:Ltot2}
\end{equation}

\subsection{LTB interior with $E=\text{const.}$}

First we calculate the action for LTB interior and $\tau = \text{const.}$ boundaries.
The Ricci scalar built from LTB metric is
\begin{equation}
{}^{(4)}\CR \sqrt{-g} = \frac{{\partial_{\rho}\left(R+R\left(\pr R\right)^2 - \left(1+2E\right)R\right)}}{\sqrt{1+2E}}\sin \theta \, .
\end{equation}
To perform the integration along the $\rho$ coordinate, we will impose the condition $E=\text{const.}$ which reduces the class of LTB models but still includes inhomogeneous matter and curvature distribution cases.
Thus, the gravitational part of LTB interior and boundary action reads
\begin{equation}
\CS_{LTB} = \frac{1}{4} \int_{t_1}^{t_2}\frac{\left(R_b\left(\pr R_b\right)^2-2ER_b\right)}{\sqrt{1+2E}}\dd \tau -\frac{1}{2} \int_{0}^{\rho_0} \frac{R^2 \pr R}{\sqrt{1+2E}}\left(\frac{\pt \pr R}{\pr R}+2\frac{\pt R_b}{R}\right)\dd \rho \, .
\end{equation}
The boundary term can be written as
\begin{equation}
\CS_{boundary} = \frac{3}{4}\int_{t_1}^{t_2} \dd \tau \frac{R_b\left(\pr R_b\right)^2}{\sqrt{1+2E}} \;\;,
\end{equation}
which leads to the total action
\begin{equation}
\CS_{grav} = -\frac{1}{2}\int^{t_2}_{t_1} \dd \tau \left(\frac{R_b\left(\partial_\rho R_b\right)^2+ER_b}{\sqrt{1+2E}}\right) \, .
\end{equation}
Changing variables to $t$ and $\psi$ gives
\begin{equation}
\CS_{grav} = -\frac{1}{2\sqrt{1+2E}}\int_{t_1}^{t_2}\dd t \sqrt{1-\frac{2M}{\psi}-\frac{\dot{\psi}^2}{1-\frac{2M}{\psi}}}\left( \frac{\dot{\psi}^2\psi}{1-\frac{2M}{\psi}-\frac{\dot{\psi}^2}{1-\frac{2M}{\psi}}}+E\psi\right)\, ,
\end{equation}
which using the hyperbolic angle can be rewritten as
\begin{equation}
\CS_{grav} = \frac{1}{2}\int_{t_1}^{t_2}\dd t \sqrt{1-\frac{2M}{\psi}}\left(\frac{-E\psi}{\cosh \mu
\sqrt{1+2E}}-\frac{\left(\psi-2M\right)}{\sqrt{1+2E}}\frac{\sinh^2\mu}{\cosh \mu}\right)\, .
  \end{equation}
The jump in the extrinsic curvature on the shell reads
\begin{equation}
[Q] = 4\sin \theta \left(\frac{\psi^2}{2\dot{f}}K_{tt}^{+} -\psi+2M+\sqrt{1+2E}\psi \dot{f} \right) \, .
\end{equation}

Lastly we consider the matter. The Lagrangian density for dust is
\begin{equation}
\CL_{dust} = -\frac{1}{2}\sqrt{-g}\epsilon \left(g_{\mu \nu}u^{\mu}u^{\nu}+1\right) \, .
\end{equation}
The action can be written as
\begin{equation}
\CS_{dust} = 2\pi \int \dd \tau \int \dd \rho \frac{R^2 \pr R}{\sqrt{1+2E}}\epsilon(\rho)({T'}^2-1)\, .
\end{equation}
In the following, we will assume that the density has an integral form given by the equation of motion \eqref{eq:eom_ltb} for LTB.
When putting $F(0)=0$ this allows us to write
\begin{equation}
\CS_{dust} = \frac{1}{2} \int \dd \tau \frac{F_b}{\sqrt{1+2E}} \left({T'}^2-1\right) \, .
  \end{equation}
Comparing with \eqref{eq:FLRW_action_dust} we see that in analogy to the FLRW case the momentum $p_T$ is given by
\begin{equation}
    p_T=\frac{F_b}{\sqrt{1+2E}}T'\, ,
\end{equation}
and as a secondary constraint we get the identification
\begin{equation}
    p_T=\frac{F_b}{\sqrt{1+2E}}\, .
\end{equation}
Recall that matching interior and exterior is done by setting $F_b=M$. To implement this here we will set $M=\sqrt{1+2E}\,p_T$ after variation.

In total this leads to the Lagrangian
\begin{multline}
	L_\text{tot} = -\frac{\psi \sqrt{1-\frac{2M}{\psi}}}{\cosh\mu}\left[\frac{1}{2\sqrt{1+2E}}\left(1-\frac{2M}{\psi} \right) \sinh^2\mu +\frac{E}{2\sqrt{1+2E}} + \frac{p_T}{\psi} - \sqrt{1+2E}\right]\\+M-\psi+\psi\dot{\psi}\mu \,. \label{eq:Ltot_LTB}
  \end{multline}
Note that this Lagrangian can be obtained from the Lagrangian for an FLRW interior \eqref{eq:Ltot} using the following identifications:
\begin{equation}
    \frac{h(\chi_b)}{C^3}\to\sqrt{1+2E}\,,\quad
    h^2(\chi_b)k\to-E\,,\quad
    \partial\chi h(\chi_b)\to\sqrt{1+2E}\,. \label{eq:identifications}
\end{equation}
It follows that the reduction undertaken for FLRW leading to \eqref{eq:Ltot2} can also be done here.
The result will be \eqref{eq:Ltot2} with the replacement \eqref{eq:identifications}.

\section{Hamiltonian formulation}\label{ham}

We can now discuss the Hamiltonian formulation for both the FLRW and LTB interior at the same time. We will work with the Lagrangian \eqref{eq:Ltot2}, keeping in mind that all results apply to the LTB model when identifying the constants involved according to \eqref{eq:identifications}. As a first step to find the corresponding Hamiltonian we want to compute the momentum conjugate to $\psi$. As $	L_\text{tot} = 	L_\text{tot}(\dot{\psi}, \mu(\dot{\psi}), M(\mu))$, we get
\begin{equation}
	p_\psi=\frac{d L_\text{tot}}{d\dot{\psi}}=\frac{\partial L_\text{tot}}{\partial \dot{\psi}} + \left( \frac{\partial L_\text{tot}}{\partial \mu} + \frac{\partial L_\text{tot}}{\partial M} \frac{\partial M}{\partial \mu} \right) \frac{\partial \mu}{\partial \dot{\psi}}\,.
\end{equation}

From \eqref{eq:Ltot2} and \eqref{eq:M} we find
\begin{align}
	\frac{\partial L_\text{tot}}{\partial \dot{\psi}} &= \psi\mu\,,\\
	\frac{\partial L_\text{tot}}{\partial \mu} &=-\psi \left(1-\frac{2M}{\psi} \right)\tanh\mu\left(\frac{C}{h^3(\chi_b)} \sqrt{1-\frac{2M}{\psi}}  \frac{\cosh^2\mu+1}{\cosh\mu}- 1\right)\,,\\
	\frac{\partial L_\text{tot}}{\partial M}&= \frac{3 C}{h^3(\chi_b)} \sqrt{1-\frac{2M}{\psi}}\frac{\sinh^2\mu}{\cosh\mu} - 1\,,\\
	\frac{\partial M}{\partial \mu} &= \psi\left(1-\frac{2M}{\psi}\right)\sinh\mu \frac{\frac{C}{h^3(\chi_b)}\sqrt{1-\frac{2M}{\psi}}\cosh\mu-1}{\frac{C}{h^3(\chi_b)}\sqrt{1-\frac{2M}{\psi}}\sinh^2\mu-\cosh\mu}\label{eq:Mmu}\,,
\end{align}
and from \eqref{eq:psidot} we find
\begin{align}
	\frac{\partial \mu}{\partial \dot{\psi}}&=\frac{\cosh^2\mu}{\left(1-\frac{2M}{\psi} \right)-\frac{2}{\psi}\frac{\partial M}{\partial\mu}\sinh\mu\cosh\mu}\\
	&= \frac{\cosh^2\mu}{1-\frac{2M}{\psi}}\frac{\frac{C}{h^3(\chi_b)}\sqrt{1-\frac{2M}{\psi}}\sinh^2\mu-\cosh\mu}{\frac{C}{h^3(\chi_b)}\sqrt{1-\frac{2M}{\psi}}\sinh^2\mu\,(1-2\cosh^2\mu)-\cosh\mu\,(1-2\sinh^2\mu)}\,.
\end{align}

The momentum is then given by
\begin{align}
	p_\psi=\psi\mu -\frac{C}{h^3(\chi_b)} \psi \sqrt{1-\frac{2M}{\psi}} \sinh\mu\,. \label{eq:ppsi}
\end{align}
Finally, the Hamiltonian can be expressed as
\begin{equation}
	H=p_\psi\dot{\psi}-L_\text{tot}= M\,, \label{eq:Hamiltonian}
\end{equation}
as was the case in \cite{KMM}.

We will now demonstrate that the Hamiltonian \eqref{eq:Hamiltonian} really gives the correct equations of motion for the Oppenheimer--Snyder model. We have to keep in mind that $M(\mu,\psi)$ is given by \eqref{eq:M} and $\mu(\psi,p_\psi)$ in turn is implicitly given by \eqref{eq:ppsi}. This gives us the equations of motion as
\begin{align}
    \dot{\psi}&=\frac{\partial M}{\partial\mu}\frac{\partial \mu}{\partial p_\psi}\,,\\
    \dot{p}_\psi&=-\frac{\partial M}{\partial\psi}-\frac{\partial M}{\partial\mu}\frac{\partial \mu}{\partial \psi}\,.
\end{align}
As it turns out, it is more useful to consider $\dot{\mu}$ instead of $\dot{p}_\psi$, which is then given by
\begin{equation}
    \dot{\mu}=-\frac{\partial M}{\partial\psi}\frac{\partial \mu}{\partial p_\psi}\,.
\end{equation}

From \eqref{eq:M} we get \eqref{eq:Mmu} as well as
\begin{equation}
    \frac{\partial M}{\partial\psi}=\frac{M}{\psi}+\frac{p_T}{\psi}\frac{\sqrt{1-\frac{2M}{\psi}}}{\frac{C}{h^3(\chi_b)} \sqrt{1-\frac{2M}{\psi}} \sinh^2\mu - \cosh\mu}\,,
\end{equation}
and from \eqref{eq:ppsi}
\begin{equation}
    \frac{\partial \mu}{\partial p_\psi}=\frac{1}{\psi} \frac{\frac{C}{h^3(\chi_b)} \sqrt{1-\frac{2M}{\psi}} \sinh^2\mu - \cosh\mu}{\frac{C}{h^3(\chi_b)} \sqrt{1-\frac{2M}{\psi}} \cosh^2\mu - \cosh\mu}\,,
\end{equation}
where we have used \eqref{eq:Mmu}.

In total, this gives us the equations of motion as
\begin{align}
    \dot{\psi}&=\left(1-\frac{2M}{\psi}\right)\tanh\mu \, ,\\
    \dot{\mu}&=-\frac{M}{\psi^2} \frac{\frac{C}{h^3(\chi_b)} \sqrt{1-\frac{2M}{\psi}} \left(\sinh^2\mu +\frac{h^3(\chi_b) p_T}{C M} \right) - \cosh\mu}{\frac{C}{h^3(\chi_b)} \sqrt{1-\frac{2M}{\psi}} \cosh^2\mu - \cosh\mu}\, . \label{eq:mudot}
\end{align}
We can see that the first equation simply gives us back the definition of the hyperbolic angle $\mu$ from the canonical formalism.

To bring the equation for $\dot\mu$ into a recognizable form, we have to do a bit more work. We want to demonstrate that the above is equivalent to the second Friedmann equation in the form \eqref{eq:Friedmann_mu}. Recall that for dust the first Friedmann equation follows directly from the second one, meaning that this is sufficient to demonstrate that our Hamiltonian gives the correct dynamics.

To this end we solve \eqref{eq:M} for $\cosh\mu$,
\begin{equation}
    \cosh\mu=\frac{1}{\sqrt{1-\frac{2M}{\psi}}}\left[ \frac{h^3(\chi_b)}{C} \pm \sqrt{\left( \partial_\chi h(\chi_b)-\frac{h^3(\chi_b)}{C}\right)^2 +\frac{2}{\psi} \left( \frac{h^3(\chi_b)}{C} p_T -M \right) } \right]\, . \label{eq:prelim}
\end{equation}
In principle, the sign should be fixed such that $\cosh\mu$ is always positive, but since it will not influence the result we will leave it open. Inserting this into the right hand side\ of \eqref{eq:mudot} allows us to integrate the equation with regard to $t$, giving
\begin{equation}
    \cosh\mu = \frac{A}{\sqrt{1-\frac{2M}{\psi}}} \left[ \frac{h^3(\chi_b)}{C} \pm \sqrt{\left( \partial_\chi h(\chi_b)-\frac{h^3(\chi_b)}{C}\right)^2 + \frac{2}{\psi} \left( \frac{h^3(\chi_b)}{C} p_T -M \right)}  \right]\,,
\end{equation}
where $A$ is a positive real constant of integration. Note that we used here that the Hamiltonian $M$ is a constant of motion. This expression is, aside from a constant, identical to \eqref{eq:prelim}, demonstrating that \eqref{eq:mudot} is simply the time derivative of \eqref{eq:prelim} and in turn \eqref{eq:M}.

Imposing $p_T = CM/h^3(\chi_b)   $ we see that \eqref{eq:prelim} gives us
\begin{equation}
    \cosh\mu=\frac{1}{\sqrt{1-\frac{2M}{\psi}}}\left[ \frac{h^3(\chi_b)}{C} \pm \left| \partial_\chi h(\chi_b)-\frac{h^3(\chi_b)}{C}\right| \right]\,, \label{eq:eom1}
\end{equation}
where the sign has to be chosen such that $\cosh\mu$ is positive. For $k=0$ this does not make a difference since there $h'(\chi_b)=h^3(\chi_b)/C=1$, giving \eqref{eq:Friedmann_mu}, using that  $h'(\chi_b)=\sqrt{1-h^2(\chi_b)k}$. For $k=\pm1$ both signs give different positive results, in addition to \eqref{eq:Friedmann_mu} also giving the solution
\begin{equation}
    \cosh\mu=\frac{2\frac{h^3(\chi_b)}{C}-\partial_\chi h(\chi_b)}{\sqrt{1-\frac{2M}{\psi}}}\,. \label{eq:friedmann_alt}
\end{equation}

In summary, we see that the Hamilton equations of motion simply give us the definition of $\mu$ (or alternatively express $p_\psi$ in terms of $\dot\psi$) and the time derivative of the equation $H=M=\text{const.}$, while this equation itself already gives the dynamics of the Oppenheimer--Snyder model, provided we choose $p_T = CM/h^3(\chi_b)   $.

It is for both $k=\pm1$ possible to identify an effective curvature $\tilde{k}$ with $0<\pm\tilde{k}<1$ as $2h^3(\chi_b)/C-\partial_\chi h(\chi_b)=\sqrt{1-h^2(\chi_b)\tilde{k}}$. One can then rescale all other quantities here, normalizing $\tilde{k}$, such that \eqref{eq:friedmann_alt} can be seen as a rescaled Friedmann equation.

For an LTB interior with $E=\text{const.}$ the situation is a bit more simple. Applying \eqref{eq:eom1} to the LTB case by using \eqref{eq:identifications} only gives the single equation of motion \eqref{eq:eom_LTB}.

\section{Conclusions}\label{con}

In this article we have developed the Hamiltonian formulation of dust collapse with the use
of a quasi-local variational principle. The  main result is that the global Hamiltonian $H$
of the OSD or LTB gravitational system, for an observer at spatial infinity, is positive definite.
This is consistent with the theorem in general relativity on positivity of the total energy of an
isolated gravitational system with non-negative local mass density (see, e.g.  \cite{RS,EW,JJ}
and references therein). Furthermore, the Hamiltonian obtained is equal to the total mass of the
system,  $H = M$, as measured by observers at infinity. This result, which resembles what is usually
obtained in Newtonian systems, albeit intuitive is not readily obtained from standard Hamiltonian
treatments in general relativity.

Our results are obtained by making use of the improved variational principle \cite{K} that was
later applied to the dynamics of a gravitational dust shell \cite{KMM}. Roughly speaking, it
consists in considering an extra boundary term added to the Hilbert action, and  assuming that
$M$, treated as a dynamical variable, is a function of an evolution parameter at the level of
the variational procedure. Then, the fact that $M$ is a constant becomes a consequence of the
equations of motion in  Hamilton's  dynamics. We can see that  the variational principle used
for the simple shell model \cite{KMM} extends to the more realistic cases of the OSD and LTB
systems, and encourages applying to still more realistic  models of collapsing  massive stars.

In the Schwarzschild black hole, $M$ is just a parameter that determines the position of the
event horizon. In our case, like a true Hamiltonian, it has an additional dynamical structure
as it depends on configuration variables and matter fields, and turns out to be constant only
as a consequence of the equations of motion.

In this article we have restricted our analysis to a single outermost dust shell of the LTB
system and we consider only the case with $E = \text{const.}$, which reduces the original
field theory LTB model  to a mechanical system. This allows to reduce the dynamics to a simple
one dimensional system where the configuration variables are position and velocity of the
outermost shell. In the homogeneous OSD case all shells obey the same dynamics as the outermost,
while in the inhomogeneous LTB case with $E= \text{const.}$ the radial position of the inner
shells can be treated as a parameter, rather than a true degree of freedom.
Due to this simplification both gravitational systems, the LTB and OSD, can be described
within one formalism. However, it should be noted that the LTB case is particularly interesting
as it may admit the existence of naked singularities (see, e.g. \cite{RG,IHD} and \cite{malafarina}),
which are appealing both from the observational and theoretical perspectives as they may
provide the valuable keys to the construction of a viable theory of quantum gravity.
More specifically, assuming that the dust density has radial dependence with a quadratic term,
i.e. $\epsilon(\rho)=\epsilon_0+\epsilon_2\rho^2$, it can be shown that there exist values for
$(\epsilon_2,E)$ for which the co-moving time of formation of the singularity coincides with the
co-moving time of formation of trapped surfaces \cite{malafarina2}.
Then, for a set of values of the boundary radius of the matter cloud, there exist null geodesics
originating at the singularity that reach far away observers  \cite{geodesics}. This seemingly
nonphysical result can be understood if one treats the singularity as the limit of a regime where
quantum gravity effects dominate. Then a naked singularity merely describes a system where quantum
gravity effects may be observable for far away observers.

Quantization of the interior of both the LTB and OSD black holes has already been done in
\cite{Kiefer:2019csi,Schmitz:2019jct}. Robustness of these results will be analyzed in the near
future. In fact, if a different quantization procedure for dust collapse were to provide the
same results this could be taken as an indication of the general validity of the results.

\acknowledgments  We would like to thank Claus Kiefer and Jerzy Kijowski for helpful discussions, and Dejan Stojkovic
for feedback on our paper. This work was partially supported by the German-Polish bilateral project DAAD and MNiSW
No 57391638, and by Nazarbayev University Faculty Development Competitive Research Grant No. 090118FD5348.


\appendix

\section{Theory of matching}\label{general}

Consider two 4-dimensional manifolds $\mathcal{M}^+$ and $\mathcal{M}^-$ separated by a 3-dimensional
time-like hypersurface $\Sigma$.
In general we may express the line element on both sides in terms of the coordinates
$\{x^0_\pm, x^1_\pm, x^2_\pm, x^3_\pm\}$ as
\be
ds^2_{\pm}=g_{\mu\nu}^\pm dx^\mu_\pm dx^\nu_\pm
\ee
while the line-element on the hypersurface, in terms of the coordinates $\{y^1,y^2,y^3\}$ is
\be
ds^2_\Sigma=\gamma_{ab}dy^a dy^b
\ee
where $a,b=1,2,3$.
The hypersurface may be expressed in parametric form on either side of the matching as
\be
\Phi^\pm(x_\pm^\mu(y^a))=0
\ee
so that the line element on the surface in terms of the coordinates on $\mathcal{M}^\pm$ is
\be
\gamma_{ab}^\pm=\frac{\partial x^\mu_\pm}{\partial y^a}\frac{\partial x^\nu_\pm}{\partial y^b}g_{\mu\nu}^\pm \, .
\ee
The induced metric is the same on both sides if we can find a set of coordinates for which
\be
\gamma^\pm_{ab}=\gamma_{ab} \, .
\ee
Let's now define the unit vector normal to $\Sigma$ on both sides as
\be
n^\pm_\mu=\frac{1}{\sqrt{|\frac{\partial \Phi^\pm}{\partial x^\alpha}\frac{\partial \Phi^\pm}{\partial
x^\beta}g^{\alpha\beta}_\pm|}}\frac{\partial \Phi^\pm}{\partial x^\mu_\pm} \, .
\ee
The second fundamental form (or extrinsic curvature) is defined by
\bea
K_{ab}^\pm&=& \frac{\partial x^\mu_\pm}{\partial y^a}\frac{\partial x^\nu_\pm}{\partial y^b}
\nabla_\mu n_\nu^\pm = \\ \nonumber
&=&-n^\pm_\sigma\left(\frac{\partial^2 x^\sigma_\pm}{\partial y^a \partial y^b}
+\Gamma^\sigma_{\mu\nu}\frac{\partial x^\mu_\pm}{\partial y^a}\frac{\partial x^\nu_\pm}{\partial y^b}\right)\, .
\eea
The boundary surface does not carry any energy-momentum tensor, and therefore the matching is smooth, if
\bea
\label{mat1}\left[\gamma_{ab}\right]&=&0 \\
\label{mat2}\left[K_{ab}\right]&=&0
\eea
where we have used the notation
\be \nonumber
\left[A\right]=A^+-A^-
\ee
for a generic quantity $A$.
 The ADM momentum density of the hypersurface is given by
\be
Q_{ab}= \sqrt{|\det \gamma_{cd} |}\left(\gamma_{ab}K- K_{ab} \right)
\ee
and its trace reads
\be
Q=\gamma^{ab}Q_{ab}=
-2\sqrt{|\det \gamma_{cd} |}\gamma^{ab}K_{ab}
\, .
\ee
For more details see \cite{EP}.


\begin{thebibliography}{99}

\bibitem{KMM} J. Kijowski, G. Magli, and D. Malafarina,
Phys. Rev. D {\bf 74}, 084017 (2006).

\bibitem{JM} P. S. Joshi and D. Malafarina,
Int. J. Mod. Phys. D, {\bf 20}, 2641 (2011).

\bibitem{HK} P. H{\'a}j{\'i}\v{c}ek and J. Kijowski,
Phys. Rev. D {\bf 62}, 044025 (2000).

\bibitem{HKA} P. H{\'a}j{\'i}\v{c}ek and C. Kiefer,
Int. J. Mod. Phys. D, {\bf 10}, 775 (2001).

\bibitem{Gozdz:2018aai} A.~ G\'{o}\'{z}d\'{z}, W.~Piechocki, and G.~Plewa,
  Eur.\ Phys.\ J.\ C {\bf 79}, 45 (2019).


\bibitem{OSD} J. R. Oppenheimer and H. Snyder,
Phys. Rev. {\bf 56}, 455 (1939);
S. Datt,
Z. Phys. {\bf 108}, 314 (1938).

\bibitem{LTB} G. Lema\^{i}tre, Ann. Soc. Sci. Bruxelles I A {\bf 53}, 51 (1933);
R. C. Tolman, Proc. Natl. Acad. Sci. USA {\bf 20}, 410 (1934);
H. Bondi, Mon. Not. Astron. Soc. {\bf 107}, 343 (1947).

\bibitem{Joshi} P. S. Joshi and I. H. Dwivedi, Commun. Math. Phys. {\bf 146}, 333 (1992).

\bibitem{K} J. Kijowski,
Gen. Rel. Grav. {\bf 29}, 307 (1997).

\bibitem{DS} A. Saini and D. Stojkovic, Phys. Rev. {\bf 89}, 044003 (2014).

\bibitem{Israel} G. Darmois, {\em Memorial de Sciences Mathematiques, Fascicule XXV Les equations de la gravitation einsteinienne}, chapter V (1927);
W. Israel, Nuvo Cemento B {\bf 44}, 1 (1966); {\bf 48}, 463 (1966).

\bibitem{matching} F. Fayos, X. Jaen, E. Llanta and J. M. M. Senovilla, Class. Quantum Grav. {\bf 8}, 2057 (1991); F. Fayos, X. Jaen, E. Llanta and J. M. M. Senovilla, Phys. Rev. D {\bf 45}, 2732 (1992); F. Fayos, J. M. M. Senovilla and R. Torres, Phys. Rev. D {\bf 54}, 4862 (1996).

\bibitem{EP} E. Poisson, {\it A Relativist's Toolkit} (Cambridge University Press, Cambridge, 2004).

\bibitem{misner} C. Misner and D. Sharp,
Phys. Rev. {\bf 136}, B571 (1964).

\bibitem{Hayward}
 G. Hayward,
 Phys. Rev. D {\bf 47}, 3275 (1993).

\bibitem{KucharBrownDust} J.~D.~Brown and K.~V.~Kucha\v{r},
Phys. Rev. D {\bf 51}, 5600 (1995).

\bibitem{RS} R. Sch\"{o}n and S. T. Yau,
Commun. Math. Phys. {\bf 65}, 45 (1979); {\bf 79}, 47 (1981); {\bf 79}, 231 (1981).

\bibitem{EW} E. Witten,  Commun. Math. Phys.
{\bf 80}, 381 (1981).

\bibitem{JJ} J. Jezierski and J. Kijowski,
Phys. Rev. D {\bf 36}, 1041 (1987).

\bibitem{RG} R. Giamb\`{o} and G. Magli, Differential Geom. Appl., {\bf 18}, 285 (2003).

\bibitem{IHD} P. S. Joshi and I. H. Dwivedi,
Phys. Rev. D {\bf 47}, 5357 (1993).

\bibitem{malafarina} D. Malafarina, Universe {\bf 3}, 48 (2017).

\bibitem{malafarina2} P. S. Joshi and D. Malafarina, IJMPD {\bf 20} 2641 (2011); P. S. Joshi, D. Malafarina and R. V. Saraykar, IJMPD {\bf 21}, 1250066 (2012).

\bibitem{geodesics} P. S. Joshi, {\em Gravitational Collapse and Spacetime Singularities}, Cambridge University Press, Cambridge, (2008).

\bibitem{Kiefer:2019csi}
  C.~Kiefer and T.~Schmitz,
  Phys.\ Rev.\ D {\bf 99}, 126010 (2019).

  \bibitem{Schmitz:2019jct}
  T.~Schmitz,
  Phys.\ Rev.\ D {\bf 101}, 026016 (2020).





\end{thebibliography}
\end{document}